\documentclass[twocolumn,showpacs,showkeys,preprintnumbers,amsmath,amssymb]{revtex4}

\usepackage{graphicx}
\usepackage{dcolumn}
\usepackage{bm}

\usepackage{tabularx}

\begin{document}

\newcommand{\br}{\mathbf{r}}
\newcommand{\bp}{\mathbf{p}}
\newcommand{\bk}{\mathbf{k}}
\newcommand{\bq}{\mathbf{q}}
\newcommand{\mh}{\mathcal{H}}
\newcommand{\mg}{\mathcal{G}}
\newcommand{\dsan}{d^3}
\newcommand{\bPi}{\boldsymbol{\Pi}}

\newcommand{\tbr}{\tilde{\mathbf{r}}}
\newcommand{\tbp}{\tilde{\mathbf{p}}}

\newcommand{\tp}{\tilde{p}}
\newcommand{\tpf}{\tilde{p}_{\text{F}}}

\newcommand{\tth}{\tilde{\theta}}

\newcommand{\tvf}{\tilde{v}_{\text{F}}}

\newcommand{\da}{$d_{x^2-y^2}$}
\newcommand{\db}{$d_{xy}$}

\newcommand{\kbets}{$\kappa\text{-(BETS)}_2\text{FeBr}_4$}
\newcommand{\lbets}{$\lambda\text{-(BETS)}_2\text{FeCl}_4$}

\preprint{APS/123-QED}

\title{
Antiferromagnetic superconductors
with effective mass anisotropy \\
in magnetic fields}
\author{Yuuichi Suginishi}
\author{Hiroshi Shimahara}%

\affiliation{
Department of Quantum Matter Science, 
ADSM, Hiroshima University, Higashi-Hiroshima 739-8530, Japan}

\date{\today}

\begin{abstract}
We derive critical field $H_{\text{c2}}$ equations 
for antiferromagnetic \textit{s}-wave, $d_{x^2-y^2}$-wave, 
and $d_{xy}$-wave superconductors 
with effective mass anisotropy in three dimensions, 
where we take into account (i)~the Jaccarino-Peter mechanism 
of magnetic-field-induced superconductivity (FISC) at high fields, 
(ii)~an extended Jaccarino-Peter mechanism that reduces 
the Pauli paramagnetic pair-breaking effect at low fields 
where superconductivity 
and an antiferromagnetic long-range order with a canted spin structure 
coexist, 
and 
(iii)~the Fulde-Ferrell-Larkin-Ovchinnikov (FFLO or LOFF) state. 
As an example, 
experimental phase diagrams observed in organic superconductor 
$\kappa\text{-(BETS)}_2\text{FeBr}_4$ 
are theoretically reproduced. 
In particular, 
the upper critical field of low-field superconductivity 
is well reproduced 
without any additional fitting parameter other than 
those determined from the critical field curves of the FISC 
at high fields. 
Therefore, 
the extended Jaccarino-Peter mechanism seems to occur actually 
in the present compound. 
It is predicted that the FFLO state does not occur in the FISC 
at high fields in contrast to the compound \lbets, 
but it may occur in low-field superconductivity 
for \textit{s}-wave and $d_{x^2-y^2}$-wave pairings.
We also briefly discuss a possibility of compounds that exhibit 
unconventional behaviors of upper critical fields. 
\end{abstract}

\maketitle

\section{\label{sec:intro}Introduction}

Recently, 
magnetic-field-induced superconductivity (FISC) 
has been observed
in organic superconductors \lbets\ 
and \kbets~\cite{uji01,balicas01,konoike04}, 
where BETS is bis(ethylenedithio)tetraselenafulvalene. 
In these salts,
localized spins on Fe$^{3+}$
exhibit antiferromagnetic long-range order 
at ambient pressure at low temperatures.
The FISC in these compounds is considered to be due to
the Jaccarino-Peter mechanism~\cite{jp62,cepas02}, 
where the localized spins are aligned uniformly at high fields.
Konoike \textit{et al.} have observed 
in the compound \kbets\ that superconductivity coexists 
with the antiferromagnetic long-range order 
in a low-field region around the zero field~\cite{konoike04}. 
They have fitted the experimental phase diagrams 
by Fisher's theory~\cite{fisher72} based on the Jaccarino-Peter mechanism.
The resultant upper critical field of low-field superconductivity 
is much smaller than their experimental data.
They have suggested that the reason for the discrepancy is that 
the Jaccarino-Peter-Fisher theory does not take into account
the antiferromagnetic long-range order at low fields~\cite{konoike04}.

In recent works, 
one of the authors has extended the Jaccarino-Peter mechanism 
to antiferromagnetic superconductors
with canted spin structures 
in magnetic fields~\cite{shimahara02a,shimahara04}.
The canted spin structure generates the ferromagnetic moments 
that create exchange fields 
acting on the conduction electrons through Kondo interactions.
If the Kondo interactions are antiferromagnetic, 
the exchange fields partly cancel the Zeeman energy.
As a result, the Pauli paramagnetic pair-breaking effect 
can be largely reduced, 
and the upper critical field can exceed 
the Pauli paramagnetic limit 
(Chandrasekhar and Clongston limit)~\cite{chandrasekhar62}.
This mechanism occurs even 
in the presence of the orbital pair-breaking effect~\cite{shimahara04}.
We call this mechanism an extended Jaccarino-Peter mechanism 
in this paper.
Since the canted antiferromagnetic phase occurs in the compound \kbets\ 
for $\mathbf{H} \parallel c$~\cite{konoike04,fujiwara01}, 
we apply the mechanism to this compound.

In the compound \kbets, 
the FISC has been observed both for 
$\mathbf{H}\parallel c$ and $\mathbf{H}\parallel a$~\cite{konoike04}.
The phase diagrams for $\mathbf{H} \parallel c$ and $\mathbf{H} \parallel a$ 
are rather different, and it
is attributed to the anisotropy of the Fermi surface 
and the Kondo interactions 
between the localized spins and the conduction electrons.
We take into account the Fermi surface anisotropy 
by effective masses.

The effective-mass model was introduced in Ginzburg-Landau equations 
by Ginzburg~\cite{ginzburg52, werthamer69}.
Hohenberg and Werthamer~\cite{hohenberg67} pointed out that
detailed structures of the Fermi surface affect the upper critical field. 
Rieck and Scharnberg~\cite{rieck90} 
and Langmann~\cite{langmann92} obtained 
general equations for arbitrary Fermi surfaces.
Xu \textit{et al.}~\cite{xu96} and 
Kim \textit{et al.}~\cite{kim98} calculated the upper critical fields 
of mixed \textit{d}-wave and \textit{s}-wave superconductors 
with effective-mass anisotropy.
Recently, Kita and Arai~\cite{kita04} have formulated an equation 
for the upper critical field, 
taking into account the Fermi surface anisotropy and the gap anisotropy 
on the basis of the Rieck and Scharnberg theory~\cite{rieck90}. 
They have performed the quantitative calculations 
of the upper critical fields for type-II superconductors 
Nb, NbSe$_2$, and MgB$_2$ using Fermi surfaces 
obtained by first-principles calculations~\cite{kita04}.

A theory of the upper critical field for layered superconductors 
has been proposed by Lebed and Yamaji~\cite{lebed98}, 
and developed by Lebed and Hayashi~\cite{lebed00}. 
They have found that when the layer spacing is large 
the upper critical field exhibits 
a reentrant transition or an enhancement 
at low temperatures in the quantum region, 
due to an effect of dimensional crossover 
induced by the magnetic field~\cite{lebed98}. 
In the compounds \kbets, however, 
since the upper critical field of the low-field superconductivity 
did not exhibit either a reentrant transition or an enhancement 
in the experimental phase diagrams, 
the dimensional crossover does not seem to take place. 
Therefore, 
from a phenomenological consideration, 
we use the effective-mass model as an approximation 
instead of strict equations in Refs.~\cite{lebed98} and \cite{lebed00}. 
The effective-mass model is adequate in the 
Ginzburg-Landau region for layered superconductors.

In this paper, 
first, we derive critical field equations 
for \textit{s}-wave, \da-wave, and \db-wave superconductors 
with effective-mass anisotropy in three directions, 
taking into account both orbital and paramagnetic pair-breaking effects. 
Secondly, we take into account the extended Jaccarino-Peter mechanism. 
Lastly, we reproduce the phase diagrams of \kbets\ 
including both the FISC and low-field superconductivity.

We also examine the possibility of the FFLO state.
The FFLO state has extensively been studied~\cite{casalbuoni04} 
since pioneering works by Fulde and Ferrell, 
and Larkin and Ovchinnikov~\cite{ff64}.
The state is taken into account by an extension of the BCS mean-field theory 
to include the possibility of finite center-of-mass momenta $\bq$'s.
In this study, 
we adopt a model in which $\bq \parallel \mathbf{H}$ is assumed 
following Gruenberg and Gunther~\cite{gruenberg66}, 
since we consider the situation in which 
substantial orbital pair-breaking effect is present.
In the organic compounds \lbets, 
the possibility of the FFLO state in the FISC has been pointed out
by Uji \textit{et al.}~\cite{uji01} 
and Balicas \textit{et al.}~\cite{balicas01}, 
and also examined theoretically~\cite{houzet02,shimahara02b}.
The shape of the phase boundary of the FISC is 
well reproduced by taking into account the FFLO state~\cite{shimahara02b}.
Tanatar \textit{et al.} have also argued 
that the FFLO state may occur 
in $\lambda\text{-(BETS)}_2\text{GaCl}_4$ 
from their experimental data~\cite{tanatar02}.
Recently, the quasi-two-dimensional heavy-fermion superconductor CeCoIn$_5$ 
has been believed to exhibit 
the FFLO state~\cite{radovan03,bianchi03,watanabe04,martin05,kakuyanagi05}.
Adachi and Ikeda~\cite{adachi03} and Won \textit{et al.}~\cite{won04} 
have calculated the critical fields for \da-wave pairing 
taking into account the FFLO state 
in connection with CeCoIn$_5$.

This paper is constructed as follows.
In the next section, 
we extend the theory of the critical field to the systems with 
anisotropic Fermi surfaces.  
The \textit{s}-wave, \da-wave, and \db-wave 
pairing superconductors are examined. 
In Sec.~\ref{sec:fisc}, we take into account 
the extended Jaccarino-Peter mechanism~\cite{shimahara02a,shimahara04}.
In Sec.~\ref{sec:res}, we apply the present theory 
to the organic superconductor \kbets.
We compare experimental and theoretical phase diagrams.
The final section is devoted to the summary and a discussion.

\section{\label{sec:form}Upper critical field 
in systems with effective-mass anisotropy}

In this section, we derive critical field equations 
for \textit{s}-wave, \da-wave, and \db-wave superconductors
with an effective-mass anisotropy.
We consider the electron system with the energy dispersion 
\begin{equation}
\label{eq:disp}
\epsilon(\bp)=\frac{p_x^2}{2m_x}+\frac{p_y^2}{2m_y}+\frac{p_z^2}{2m_z} 
=\sum_{\nu=1}^3\frac{p_\nu^2}{2m_\nu} 
\end{equation}
and the pairing interaction 
\begin{equation}
\label{eq:inth}
\mh' = 
\int \dsan\br \int \dsan\br'\psi_\uparrow^\dag(\br)\psi_\uparrow(\br)
V(\br-\br')\psi_\downarrow^\dag(\br')\psi_\downarrow(\br') \;,
\end{equation}
where we have defined 
$\bp=(p_1, p_2, p_3)=(p_x, p_y, p_z)$
and introduced anisotropic effective masses 
$m_x=m_1$, $m_y=m_2$, and $m_z=m_3$.
In the magnetic field $\mathbf{H}=\mathbf{B}= \text{rot}\mathbf{A}$, 
the Hamiltonian is written as
\begin{equation}
\mh = \mh_0 + \mh_m + \mh' \;,
\end{equation}
where
\begin{equation}
\label{eq:allh}
\begin{split}
\mh_0 &= 
\sum_\sigma \int \dsan\br \psi_\sigma^\dag(\br) 
\\ & \quad \times
\sum_{\nu = 1,2,3} \frac{1}{2m_\nu}
 \left[ -i \hbar \frac{\partial}{\partial x_\nu} 
 - \frac{e}{c}A_\nu(\br)\right]^2
 \psi_\sigma(\br) \;, \\
\mh_m &= 
 -\int \dsan\br \; \mu_{\text{e}}\mathbf{H} \cdot \left[
 \sum_{\sigma\sigma'}\psi_\sigma^\dag(\br) 
 \boldsymbol{\sigma}_{\sigma\sigma'} 
 \psi_{\sigma'}(\br) \right] \;.
\end{split}
\end{equation}
Here, $\psi_\sigma(\br)$ 
denotes the field operator 
which annihilates an electron of spin $\sigma$ at a point 
$\br\equiv(x,y,z)\equiv(x_1,x_2,x_3)$.
We have defined the electronic magnetic moment 
$\mu_{\text{e}}=-g_{\text{e}}\mu_{\text{B}}/2$ 
with the Bohr magneton $\mu_{\text{B}}=\hbar|e|/(2mc)$ 
and the \textit{g} factor of the conduction electrons $g_{\text{e}}$. 
We use the units such that $c=k_{\text{B}}=\hbar=1$ 
unless it is explicitly expressed.

We define the $S^z$ axis in the spin space along the direction of $\mathbf{H}$.
We should note that the $S^z$ axis does not necessarily 
coincide with the $z$ axis of the electron coordinate 
depending on the direction of the $\mathbf{H}$.
Therefore, we have 
\begin{equation}
\mh_m=-\sum_\sigma \int \dsan\br \; \sigma h 
  \psi_\sigma^\dag(\br)\psi_\sigma(\br) \;,
\end{equation}
where $h \equiv \mu_{\text{e}} H$ with $H = |\mathbf{H}|$, 
and $\sigma=+1$ and $-1$ for up and down spin states, respectively.

We apply the BCS mean-field approximation 
to the interaction Hamiltonian~(\ref{eq:inth}) as
\begin{equation}
\begin{split}
\mh' &= 
-\int \dsan\br \int \dsan\br' \Bigl[
 \Delta_\downarrow(\br,\br')
 \psi_\uparrow^\dag(\br)\psi_\downarrow^\dag(\br') \\
& \qquad \qquad \quad \quad \quad \;
 +\Delta_\downarrow^\ast(\br,\br')
 \psi_\downarrow(\br')\psi_\uparrow(\br) \Bigl] \;,
\end{split}
\end{equation}
defining the order parameter
\begin{equation}
\Delta_{-\sigma}(\br,\br') \equiv -V(\br-\br') 
 \langle \psi_{-\sigma}(\br')\psi_\sigma(\br) \rangle \;.
\end{equation}
We define the center-of-mass coordinate $\mathbf{R}=(\br + \br')/2$ 
and the relative coordinate $\boldsymbol{\rho}=\br-\br'$, 
and redefine the gap function as $\Delta_{-\sigma}(\mathbf{R},\boldsymbol{\rho})$.
By following the procedure of Refs.~\cite{scharnberg80} and \cite{shimahara96}, 
we obtain the linearized gap equation 
\begin{equation}
\label{eq:gap1}
\begin{split}
\Delta&_{-\sigma}(\br,\bp)
\\ 
&=-T\sum_n \int\frac{\dsan\bp'}{(2\pi)^3}
 \int\dsan\boldsymbol{\rho} \; \exp(i\bp'\cdot\boldsymbol{\rho}) 
\\ & \qquad \qquad \quad \times
 V(\bp-\bp') G_\sigma^{(0)}(-\bp',-i\omega_n) 
\\ & \qquad \qquad \times
 \int\frac{\dsan\bp''}{(2\pi)^3} \; \exp(i\bp''\cdot\boldsymbol{\rho})
 G_{-\sigma}^{(0)}(\bp'',-i\omega_n)
\\ & \qquad \qquad \quad \times
 \exp(i\boldsymbol{\rho}\cdot\bPi)
 \Delta_{-\sigma}(\br,\bp') 
\end{split}
\end{equation}
near the second-order phase-transition point. 
We could examine both orbital and paramagnetic pair-breaking effects 
with Eq.~(\ref{eq:gap1}).
In the derivation of Eq.~(\ref{eq:gap1}), 
it is assumed that the spatial variation of vector potential $\mathbf{A}$ 
is sufficiently slow.
Therefore, 
the order parameter becomes a slowly varying function of 
the center-of-mass coordinate.
We have defined 
\begin{equation}
\label{eq:pi}
\bPi=-i\hbar\nabla-\frac{2e}{c}\mathbf{A}(\br) \;,
\end{equation}
the Matsubara frequencies $\omega_n=(2n+1)\pi T$ 
with integer $n=0, \pm 1, \pm2, \dots$, 
and the free-electron Green's function 
\begin{equation}
\label{eq:freeg}
G_\sigma^{(0)}(\bp,i\omega_n)=\frac{1}{i\omega_n-\epsilon(\bp)+\sigma h + \mu} \;.
\end{equation}

We introduce 
$\tbp=(\tilde{p}_x,\tilde{p}_y,\tilde{p}_z)=(\tilde{p}_1,\tilde{p}_2,\tilde{p}_3)$ 
with $\tilde{p}_\nu=(\tilde{m}/m_\nu)^{1/2}p_\nu$ 
so that the dispersion relation becomes isotropic as
\begin{equation}
\epsilon(\bp)=\tilde{\epsilon}(\tbp) 
 \equiv \frac{\tilde{p}^2}{2\tilde{m}}
 =\frac{1}{2\tilde{m}} \left( \tilde{p}_x^2+\tilde{p}_y^2+\tilde{p}_z^2 \right) 
\end{equation}
in the $\tilde{\bp}$ space, 
where $\tilde{m} \equiv (m_x m_y m_z)^{1/3}$.
We also introduce 
$\tbr=(\tilde{x}_1, \tilde{x}_2, \tilde{x}_3) 
= (\tilde{x}, \tilde{y}, \tilde{z})$ 
with $\tilde{x}_\nu=(m_{\nu}/\tilde{m})^{1/2}x_\nu$ 
and $\partial / \partial \tilde{x}_\nu 
=(\tilde{m}/m_\nu)^{1/2}\partial / \partial x_\nu$, 
so that $\bp\cdot\br = \tbp\cdot\tbr$.
We also define the operator $\tilde{\bPi}$ with
\begin{equation}
\label{eq:tpi}
\tilde{\boldsymbol{\rho}}\cdot\tilde{\bPi} 
\equiv \boldsymbol{\rho}\cdot\bPi \;, 
\end{equation}
where $\boldsymbol{\rho}=(\rho_1,\rho_2,\rho_3)$ and
$\tilde{\rho}_\nu = (m_\nu/\tilde{m})^{1/2}\rho_\nu$.
We will calculate the explicit form of 
$\tilde{\bPi}$ afterward.

We replace $|\tilde{\bp}|$ and $|\tilde{\bp}'|$ 
in $V(\bp-\bp')$, 
$\Delta_{-\sigma}(\br,\bp)$, 
and $\Delta_{-\sigma}(\br,\bp')$
with the Fermi momentum $\tilde{p}_{\text{F}}$ 
since electrons with momenta $\tilde{\bp}$ and $\tilde{\bp}'$ 
near the Fermi surface mainly contribute to the gap equation.
Therefore, we write $V(\bp-\bp')$ and $\Delta_{-\sigma}(\br,\bp)$ 
as $V(\hat{\bp},\hat{\bp}')$ and $\Delta_{-\sigma}(\tbr,\hat{\tbp})$, 
respectively, 
with unit vectors $\hat{\bp}=\bp/|\bp|$ 
and $\hat{\tilde{\bp}}=\tilde{\bp}/|\tilde{\bp}|$.

After simple algebra 
using Eq.~(\ref{eq:freeg}), 
we obtain
\begin{equation}
\label{eq:gap2}
\begin{split}
\Delta&_{-\sigma}(\tbr,\hat{\tbp})
\\ &
 =-\pi T \sum_{|\omega_n|<\omega_{\text{D}}} N(0) 
 \int\frac{d\Omega_{\hat{\tbp}'}}{4\pi}
 V(\hat{\bp},\hat{\bp}') 
\\ & \qquad\qquad \times
 \int_0^\infty \!\!\! dt \;
 e^{-[|\omega_n|-i\sigma h \; \text{sgn}(\omega_n)]t}
\\ & \qquad\qquad \times 
 \exp\left[\frac{t}{2i}\text{sgn}(\omega_n)
 \tilde{\mathbf{v}}_{\text{F}}(\hat{\tbp}')
 \cdot\tilde{\bPi} \right]
 \Delta_{-\sigma}(\tbr,\hat{\tbp}') \;,
\end{split}
\end{equation}
where we have defined 
$\tilde{\mathbf{v}}_{\text{F}}(\tbp) \equiv \tbp/\tilde{m}$ and 
the density of states at the Fermi energy 
$N(0) \equiv \tilde{m} 
\tilde{p}_{\text{F}}/(2 \pi^2 \hbar^2)$~\cite{shimahara96}.
Since Eq.~(\ref{eq:gap2}) does not depend 
on the spin value $\sigma$, 
we omit the spin suffix from now on.
By redefining 
$\tbp'=t\tilde{\mathbf{v}}_{\text{F}}(\hat{\tbp}')/2
=t\tvf\hat{\tbp}'/2$ 
and summing up the Matsubara frequencies, 
the gap equation (\ref{eq:gap2}) acquires the form
\begin{equation}
\label{eq:gap3}
\begin{split}
\Delta(\tbr,\hat{\tbp}) = -\frac{TN(0)}{2\tvf}
 \int&\dsan\tbp' V(\hat{\bp},\hat{\bp}') 
 \frac{1-e^{-2\tilde{p}'\omega_{\text{D}}/\tvf}}
 {\mbox{${\tilde p}'$}^2 \sinh(2\pi T \tilde{p}' / \tvf)}
\\ & \times
 \cos\left[\frac{2\tilde{p}'}{\tvf}h 
 + \tbp'\cdot\tilde{\boldsymbol{\Pi}} 	\right]
 \Delta(\tbr,\hat{\tbp}') \;.
\end{split}
\end{equation}
It is obvious that if we rewrite the integral variable $\tbp$ as $\bp$, 
the linearized gap equation~(\ref{eq:gap3}) 
is similar to those 
obtained by many authors~\cite{werthamer69, hohenberg67, 
luk'yanchuk87, scharnberg80, rieck90, shimahara96, shimahara97a,prohammer90}.
The differences between Eq.~(\ref{eq:gap3}) 
and the equations obtained so far for isotropic systems 
are that the vector potential $\mathbf{A}$ 
is scaled as $\tilde{A}_\nu=(m_\nu /\tilde{m})^{1/2}A_\nu$ 
in $\tilde{\boldsymbol{\Pi}}$, 
and that the pairing interaction $V(\hat{\bp},\hat{\bp}')$ 
is deformed in the $\tbp$-space anisotropic superconductivity.
The former scale transformation in $\mathbf{A}$ 
has been studied in Ginzburg-Landau theory~\cite{ginzburg52,werthamer69,klemm80} 
for \textit{s}-wave pairing. 
For \textit{s}-wave pairing, 
since $V(\hat{\bp},\hat{\bp}') = V(\hat{\tbp},\hat{\tbp}')$, 
Eq.~(\ref{eq:gap3}) is exactly reduced to the equations 
for the system with the isotropic Fermi surface, 
except the scale of $A_\nu$, 
which results in the scaling $\tilde{H}_\nu=(m_\nu/\tilde{m})^{1/2}H_\nu$.
Therefore, 
the only difference due to the Fermi surface anisotropy 
is that the critical field in the $\nu$ direction 
is enhanced or reduced by the factor $(\tilde{m}/m_\nu)^{1/2}$.
The latter deformation of $V(\hat{\bp},\hat{\bp}')$ 
occurs because $V(\hat{\bp},\hat{\bp}') \ne V(\hat{\tbp},\hat{\tbp}')$ 
for anisotropic superconductivity.
We take into account the deformation of the pairing interaction 
in the $\tilde{\bp}$ space.
These changes affect 
the mass anisotropy dependence of the critical field.

Now, we derive a more explicit form of the upper critical field equation.
We take the \textit{z} axis in the direction 
where the effective mass is the largest.
The compounds with layered structures 
with large layer spacing 
are approximately described by the models with $m_z \gg m_x, m_y$.
For example, in the application to the organic superconductor \kbets, 
we take the $x$ and $y$ axes along $c$  and $a$ axes of the compounds, respectively. 
We study a quasi-two-dimensional superconductor under magnetic fields 
parallel to the layers 
($\mathbf{H} \parallel x$ or $\mathbf{H} \parallel y$) hereafter.
We derive the critical field equations only for $\mathbf{H} \parallel x$, 
because those for $\mathbf{H} \parallel y$ 
are obtained by exchanging $m_x$ and $m_y$.

We assume the uniform magnetic field $\mathbf{H}$ 
in the $-x$ direction parallel to the layers, 
which is expressed by a vector potential $\mathbf{A}=(0,0,-Hy)$.
Then, we obtain the explicit form of $\tilde{\boldsymbol{\Pi}}$ as 
\[
\tilde{\bPi}
 = \left(-i\hbar\frac{\partial}{\partial \tilde{x}},\;
    -i\hbar\frac{\partial}{\partial \tilde{y}}, \;
    -i\hbar\frac{\partial}{\partial \tilde{z}}
    -\frac{2|e|}{c}\frac{\tilde{m}}{\sqrt{m_y m_z}}H\tilde{y} \right) 
\]
from Eqs.~(\ref{eq:pi}) and (\ref{eq:tpi}).
We define the differential operators 
\begin{equation}
\begin{split}
\eta_{\tbr}&=\frac{1}{\sqrt{2\kappa_x}} 
 (\tilde{\Pi}_y-i\tilde{\Pi}_z) 
\;,\\
\eta_{\tbr}^\dag&=\frac{1}{\sqrt{2\kappa_x}} 
 (\tilde{\Pi}_y+i\tilde{\Pi}_z) 
\;,
\end{split}
\end{equation}
which satisfy the bosonic commutation relations, where 
\begin{equation}
\kappa_x \equiv \frac{\tilde{m}}{\sqrt{m_y m_z}}\frac{2|e|H}{c} \;.
\end{equation}
The factor $\tilde{m}/\sqrt{m_ym_z}$ is due to 
the effective-mass anisotropy.

We consider pairing interactions of the form 
\begin{equation}
\label{eq:gpi}
V(\hat{\bp},\hat{\bp}') 
= -g_\alpha \gamma_\alpha(\hat{\bp})\gamma_\alpha(\hat{\bp}') \;, 
\end{equation}
where $\gamma_\alpha (\hat{\bp})$ denotes the symmetry function 
of symmetry $\alpha$ 
and $\alpha=s$, \da, \db\, and so on.
The symmetry function $\gamma_\alpha(\hat{\bp})$ is normalized so that 
$ \langle[\gamma_{\alpha}(\hat{\bp})]^2\rangle = 1 $, 
where the average on the Fermi surface $\langle\cdots\rangle$ is defined by
$\langle\cdots\rangle 
 =  \int d\Omega_\bp \rho(0,\theta,\varphi)(\cdots) /
    \int d\Omega_\bp \rho(0,\theta,\varphi)$, 
where $\rho(0,\theta,\varphi)$ 
denotes the angle-dependent density of states on the Fermi surface.
The gap function is proportional to 
the symmetry function $\gamma_\alpha(\hat{\bp})$ 
and expanded by the Abrikosov functions $\phi_n(\tbr)$ 
of the Landau-level indexes $n=0,1,2,\dots$, as 
\begin{equation}
\label{eq:gf1}
\Delta(\tbr,\hat{\tbp}) 
 = \sum_{n=0}^{\infty} \Delta_n^\alpha \gamma_\alpha(\hat{\bp}) 
 \phi_n(\tbr)e^{i\tilde{q}\tilde{x}} \;.
\end{equation}
The Abrikosov functions $\phi_n(\tbr)$ are expressed as 
\begin{equation}
\phi_n(\tbr)=\frac{1}{\sqrt{n!}}(\eta_{\tbr}^\dag)^n\phi_0(\tbr) \;,
\end{equation}
where $\phi_0(\tbr)$ denotes the solution of $\eta_{\tbr}\phi_0(\tbr)=0$. 
In the gap function~(\ref{eq:gf1}), we have taken into account the possibility 
of nonzero $\tilde{q}$ for the FFLO state, 
where ${\tilde q} \equiv ({\tilde m}/{m_x})^{1/2} |\bq|$. 
The linearized gap equation~(\ref{eq:gap3})
can be written as
\begin{equation}
\label{eq:mgeq1}
\Delta_n^\alpha=N(0)g_\alpha \sum_{n'}E_{nn'}^\alpha\Delta_{n'}^\alpha 
\end{equation}
with
\begin{equation}
\label{eq:mgeq2}
E_{nn'}^\alpha 
 = \delta_{nn'} E_0^{\alpha} - D_{nn'}^\alpha
\;, 
\end{equation}
where we have defined 
\begin{equation}
\label{eq:E0Dnndef}
\begin{split}
E_0^{\alpha}
 & 
 = 
 \frac{T}{2\tvf} 
 \int \dsan\tbp [\gamma_\alpha(\hat{\bp})]^2 
 \frac{1-e^{2\tilde{p} \omega_{\text{D}}/\tvf}}
      {\tilde{p}^2 \sinh(2 \pi T \tilde{p} / \tvf)} \;,
\\
D_{nn'}^\alpha 
&=-\frac{T}{2\tvf} 
 \int \dsan\tbp [\gamma_\alpha(\hat{\bp})]^2 
 \frac{1-e^{2\tilde{p} \omega_{\text{D}}/\tvf}}
      {\tilde{p}^2 \sinh(2 \pi T \tilde{p} / \tvf)}
 e^{i(n'-n)\tilde{\varphi}}
\\ &\qquad\qquad\times
 \Bigg[
 \sum_{j=1}^\infty 
 d_{nn'}^{(j)}(\tilde{p},\tth)
 \left(\delta_{n'-n,2j} + \delta_{n'-n,-2j}\right) 
\\ & \qquad\qquad\qquad\qquad
 + d_{nn'}^{(0)}(\tilde{p},\tth)\delta_{nn'} 
 \Bigg]
\end{split}
\end{equation}
and 
\begin{equation}
\begin{split}
d_{nn'}^{(j)}(\tilde{p},\tth) 
 \equiv &\exp\left[-\frac{\kappa_x}{4}\tilde{p}^2\sin^2\tth\right]
\\ & \times
 \sum_{k=0}^{\min(n,n')}
 \left[ -\frac{\kappa_x}{2} \tilde{p}^2\sin^2\tth \right]^{k+j}
\\ & \qquad \times
 \frac{\sqrt{n!n'!}}{k!(k+2j)![\min(n,n')-k]!} 
\\ & \qquad \times
 \cos\left[\frac{2\tilde{p}}{\tvf}
 (h+\frac{\tilde{q}\tvf}{2}\cos\tth)\right]
 -\delta_{nn'}\delta_{j,0} \;.
\end{split}
\end{equation}
More explicit expressions of the matrix elements 
$D_{nn'}^{s}$, $D_{nn'}^{d_{x^2-y^2}}$, and $D_{nn'}^{d_{xy}}$ 
are given below. 
Defining the zero-field transition temperature $T_{\text{c}}^{(0)}$, 
the linearized gap equation~(\ref{eq:mgeq1}) is written as 
\begin{equation}
\label{eq:mgeqg}
\begin{split}
\log\frac{T}{T_{\text{c}}^{(0)}}\Delta_{n}^\alpha
 =-\sum_{n'=0}^\infty D_{nn'}^\alpha
 \Delta_{n'}^\alpha \;.
\end{split}
\end{equation}
The transition temperature and the critical field are given by 
the condition that Eq.~(\ref{eq:mgeq1}) has 
a nontrivial solution of $\Delta_n^\alpha$ for the first time  
when the magnetic field and temperature decrease, respectively, 
where $\tilde{q}$ is optimized.

\subsection{The case of \textit{s}-wave pairing}

For \textit{s}-wave pairing, 
we insert $\gamma_s(\hat{\bp})=1$ into Eq.~(\ref{eq:mgeq2}).
It is easily verified by integrating over $\tilde{\varphi}$
that $E_{nn'}^s=0$ for $n \ne n'$.
Therefore, the matrix elements $D_{nn'}^s$
in the linearized gap equation ~(\ref{eq:mgeqg}) are expressed as
\begin{equation}
\label{eq:mgeqs}
D_{nn'}^s = -\frac{\pi T}{\tvf}
 \int_0^\infty \!\! d\tilde{p} \int_0^\pi \!\! \sin \tth d\tth
 \frac{d_{nn'}^{(0)}(\tilde{p},\tth) \delta_{nn'}}{\sinh(2\pi T\tilde{p}/\tvf)} \;.
\end{equation}

\subsection{The case of \da-wave pairing}

We consider \da-wave pairing interaction 
given by Eq.~(\ref{eq:gpi}) with 
$ \gamma_{d_{x^2-y^2}}(\hat{\bp}) = C_1(\hat{p}_x^2-\hat{p}_y^2) $, 
where $(\hat{p}_x, \hat{p}_y)=(p_x, p_y)/\sqrt{p_x^2+p_y^2}$.
The function $\gamma_{d_{x^2-y^2}}(\hat{\bp})$ is rewritten as
\begin{equation}
\label{eq:gda}
\begin{split}
\gamma_{d_{x^2-y^2}}(\hat{\bp})
 &=C_1\frac{m_x\tilde{p}_x^2-m_y\tilde{p}_y^2}{m_x\tilde{p}_x^2+m_y\tilde{p}_y^2} 
\\ &
 =C_1\left[ \frac{2m_x}{m_x+m_y\tan^2\tth\cos^2\tilde{\varphi}}
 -1\right] \;,
\end{split}
\end{equation}
where we have defined the polar coordinates ($\tth, \tilde{\varphi}$) 
taking the $\tilde{p}_x$ axis as the polar axis, so that
$\tth$ is measured from the $\tilde{p}_x$ axis, 
and $\tilde{\varphi}$ is measured from the $\tilde{p}_y$ axis. 
Therefore, in the linearized gap equation~(\ref{eq:mgeqg}), 
the matrix elements $D_{nn'}^{d_{x^2-y^2}}$ 
are written as 
\begin{equation}
\label{eq:mgeqd1}
\begin{split}
D_{nn'}^{d_{x^2-y^2}} 
 =D&_{nn'}^{d_{x^2-y^2}(0)}\delta_{nn'} 
\\ &
 + \sum_{j=1}^\infty 
 D_{nn'}^{d_{x^2-y^2}(j)} (\delta_{n,n'+2j}+\delta_{n+2j,n'}) \;,
\end{split}
\end{equation}
where $D_{nn'}^{d_{x^2-y^2}(j)}$ is defined by 
\begin{equation}
\begin{split}
D_{nn'}^{d_{x^2-y^2}(j)} = - &\frac{\pi T}{\tvf}
 \int_0^\infty \!\! d\tilde{p} 
\\ & \times
 \int_0^\pi \!\! \sin\tth d\tth
 \frac{C_{d_{x^2-y^2}}^{(j)}(\tth) d_{nn'}^{(j)}(\tilde{p},\tth)}{\sinh(2\pi T\tilde{p}/\tvf)} 
 \;,
\end{split}
\end{equation}
with
\begin{equation}
\begin{split}
C_{d_{x^2-y^2}}^{(j)}(\tth) 
 = & \left[ 1+\frac{2\sqrt{m_xm_y}}{m_x+m_y} \right]
 \left[\frac{c(\tth)-1}{c(\tth)+1} \right]^j
\\ & \times
 \Bigg{\{} \delta_{j,0}-2c(\tth)
 +4j[c(\tth)]^2
\\ & \qquad
 +2[c(\tth)]^3\Bigg{\}} \;,
\end{split}
\end{equation}
and $c(\tth) \equiv 1/\sqrt{1+(m_y/m_x)\tan^2 \tth}$.
The matrix equation~(\ref{eq:mgeqg}) 
is decoupled two sets of equations 
for $\Delta_n^{d_{x^2-y^2}}$ of even and odd $n$'s.

\subsection{The case of \db-wave pairing}

Next, 
we consider \db-wave pairing interaction with
\begin{equation}
\gamma_{d_{xy}}(\hat{\bp}) 
 = C_2\hat{p}_x\hat{p}_y 
 = C_2
 \frac{\sqrt{m_xm_y}\tan\tth\cos\tilde{\varphi}}
      {m_x+m_y\tan^2\tth\cos^2\tilde{\varphi}} \;.
\end{equation}
%
%
The matrix elements $D_{nn'}^{d_{xy}}$ take the form 
\begin{equation}
\label{eq:mgeqd2}
\begin{split}
D_{nn'}^{d_{xy}} 
 = D&_{nn'}^{d_{xy}(0)} \delta_{nn'} 
\\ &
 + \sum_{j=1}^\infty 
 D_{nn'}^{d_{xy}(j)} (\delta_{n,n'+2j}+\delta_{n+2j,n'}) \;,
\end{split}
\end{equation}
where $D_{nn'}^{d_{xy}(j)}$ is defined by 
\begin{equation}
D_{nn'}^{d_{xy}(j)} = - \frac{\pi T}{\tvf}
 \int_0^\infty \!\! d\tilde{p} \int_0^\pi \!\! \sin\tth d\tth
 \frac{C_{d_{xy}}^{(j)}(\tth) d_{nn'}^{(j)}(\tilde{p},\tth)}{\sinh(2\pi T\tilde{p}/\tvf)} \;,
\end{equation}
with
\begin{equation}
\begin{split}
C_{d_{xy}}^{(j)}(\tth) 
 = & \left[ 1+\frac{m_x+m_y}{2\sqrt{m_xm_y}} \right]
\\ & \times
 \Bigg{\{} 2c(\tth)\left\{1-[c(\tth)]^2\right\} \delta_{j,0}
\\ & \qquad\quad
 -\frac{4j}{[c(\tth)]^{2j-1}}
 \left[\frac{c(\tth)-1}{c(\tth)+1} \right]^j
 \Bigg{\}} \;.
\end{split}
\end{equation}
The order parameter components 
$\Delta_0^{d_{xy}}$ and $\Delta_1^{d_{xy}}$ couple only 
with $\Delta_{2n}^{d_{xy}}$ and $\Delta_{2n+1}^{d_{xy}}$, 
respectively, where $n=1,2,\dots$ .
Therefore the matrix equation~(\ref{eq:mgeqg}) 
for \db-wave pairing is decoupled
into two sets of equations for $\Delta_n^{d_{xy}}$ of even and odd $n$'s.

\section{\label{sec:fisc}Extended Jaccarino-Peter mechanism}

In this section, 
we review an extended Jaccarino-Peter mechanism 
in antiferromagnetic superconductors~\cite{shimahara02a,shimahara04}.
We have formulated the upper critical field equation of
the conduction electron system in the previous section.
Now, we shall introduce a localized spin system 
that is coupled with the conduction electron system 
through the  Kondo interaction $J_{\text{K}}$.
We consider a situation in which 
the antiferromagnetic long-range order 
has canted spin structures in magnetic fields.
Then, ferromagnetic moments 
are induced in the localized spins 
and create exchange fields on the conduction electrons.
The exchange fields partly or completely 
cancel the magnetic fields 
in the Zeeman energy terms of the conduction electrons 
when $J_{\text{K}} > 0$.
The exchange fields are not real magnetic fields in the sense 
that they only modify the Zeeman energy terms
but not the vector potential.
On the other hand, 
the ferromagnetic moments also create real internal magnetic fields.
However, such internal magnetic fields are weak 
when the localized spins are spatially separated 
from the conduction electrons.
Therefore, we neglect them in this study.

We examine the case in which the antiferromagnetic transition 
occurs at a temperature much higher 
than the superconducting transition temperature 
and thus the magnetic ordered state is rigid. 
Therefore, we neglect modification of the magnetic structure 
by occurrence of superconductivity. 
Strictly speaking, 
the localized spin state is modified by the mobile electrons, 
and the state of the total system 
should be determined self-consistently~\cite{hamada95}.
However, 
we consider a model in which 
such modification has already been included 
and the canted spin structure occurs as a result.

We modify the extended Jaccarino-Peter mechanism~\cite{shimahara02a,shimahara04} 
for the system with anisotropic Kondo interactions, 
introducing Kondo coupling constants $J_{\text{K}}^\mu$ 
between the localized spins and 
the conduction electrons of the $x_\mu$ component in spin space.
When the magnetic field is oriented to the $-x_\mu$ direction, 
where $\mu = 1, 2, 3$, 
we replace $\mh_m$ in Eq.~(\ref{eq:allh}) with
the effective Hamiltonian 
\begin{equation}
\tilde{\mh}_m = 
-\sum_{\sigma} \int \dsan\br \;
 \sigma \tilde{h}_\mu
 \psi_\sigma^\dag(\br) 
 \psi_{\sigma}(\br) \;,
\end{equation}
where $\tilde{h}_\mu$ denotes the effective Zeeman energy. 
Therefore, in the critical field equations derived in the previous section, 
the Zeeman energy $h$ is replaced with $\tilde{h}_\mu$.
For $H \le H_{\text{AF}}$, 
$\tilde{h}_\mu$ is given by 
\begin{equation}
\label{eq:thlf}
\tilde{h}_\mu
 =\left(1-
 \frac{g_{\text{s}}z_{\text{K}}J_{\text{K}}^\mu}{g_{\text{e}}zJ}\right)h
 =\left(1-\frac{J_{\text{K}}^\mu}{J_{\text{AF}}'}\right)h 
\end{equation}
with $J_{\text{AF}}' \equiv g_{\text{e}}zJ/(g_{\text{s}}z_{\text{K}})$.
In Eq.~(\ref{eq:thlf}), $z$,  
$z_{\text{K}}$, $J$, and $g_{\text{s}}$
denote number of antiferromagnetic bonds for a given site,
number of lattice sites 
that participate in the Kondo interaction 
with the conduction electrons at $\br$, 
coupling constant of the exchange interaction of the localized spins, 
and the \textit{g} factor of the localized spins, respectively.
Here, $H_{\text{AF}}$ denotes the critical field of the antiferromagnetic phase 
which is given by $H_{\text{AF}} = 2zJ\bar{S}/g_{\text{s}}\mu_{\text{B}}$, 
where $\bar{S}$ denotes the magnitude of the localized spins.
Therefore, when $0 < J_{\text{K}}^\mu < 2J_{\text{AF}}'$, 
the Pauli paramagnetic pair-breaking effect is reduced 
in antiferromagnetic superconductors.

For $H \ge H_{\text{AF}}$, 
since the spins are aligned uniformly, 
the effective Zeeman energy $\tilde{h}_\mu$ is given by
\begin{equation}
\tilde{h}_\mu=h-z_{\text{K}}J_{\text{K}}^\mu\bar{S} \;.
\end{equation}
Because the second term is a negative constant when $J_{\text{K}}^\mu>0$, 
the present mechanism is reduced to the conventional 
Jaccarino-Peter mechanism.
The Zeeman energy is completely compensated 
at $H = z_{\text{K}}J_{\text{K}}^\mu\bar{S}/|\mu_{\text{e}}| 
\equiv H_{\text{cent}}^\mu$.
It is noted that FISC occurs only when 
$H_{\text{cent}}^\mu / H_{\text{AF}} 
 = J_{\text{K}}^\mu / J_{\text{AF}}' \gtrsim 1$~\cite{shimahara02a,shimahara04}.

\section{\label{sec:res}Application to an organic superconductor}

\begin{figure}[t]
\begin{center}
\includegraphics[width=65mm,keepaspectratio]{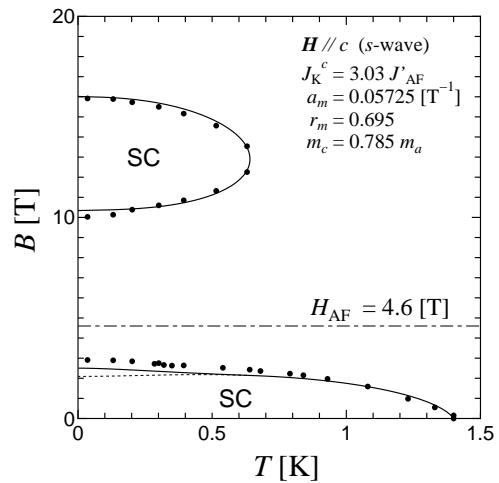}
\caption{\label{fig:sc} 
Theoretical phase diagram for \textit{s}-wave pairing
in the magnetic fields parallel to the \textit{c} axis.
Here, SC stands for superconductivity. 
The solid and dotted curves show boundaries of superconductivity
in the presence and the absence of the FFLO state, respectively.
The closed circles show the experimental data 
of the superconducting transition points 
for $\mathbf{H} \parallel c$ in \kbets\ 
by Konoike \textit{et al}.~(Ref.~\cite{konoike04}).
The dot-dashed line shows the critical field 
of the antiferromagnetic phase.}
\end{center}
\end{figure}

\begin{figure}[htbp]
\begin{center}
\includegraphics[width=65mm,keepaspectratio]{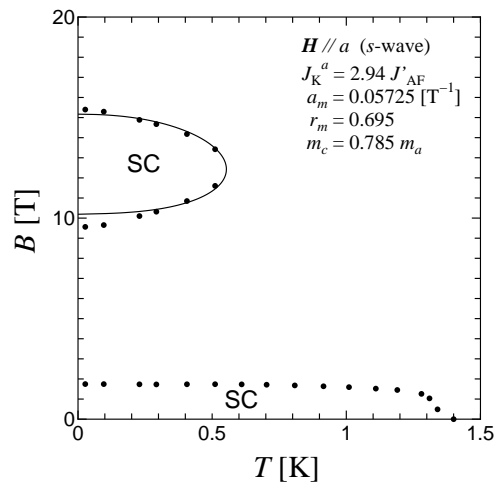}
\caption{\label{fig:sa} 
Theoretical phase diagram for \textit{s}-wave pairing
in the magnetic fields parallel to the \textit{a} axis.
The notations are the same as those in Fig.~\ref{fig:sc}.
Since the canted spin structure does not occur 
for $\mathbf{H} \parallel a$ at low field, 
the upper critical field of low-field superconductivity 
has not been calculated.}
\end{center}
\end{figure}

In this section, 
we apply the present theory to 
the antiferromagnetic organic superconductor \kbets\ 
taking the \textit{x} and \textit{y} axes along 
the crystallographic \textit{c} and \textit{a} axes, respectively. 
Therefore, we write the effective masses as 
$m_x=m_c$, $m_y=m_a$, and $m_z=m_b$,  
and the anisotropic Kondo interactions as 
$J_{\text{K}}^1=J_{\text{K}}^c$ and $J_{\text{K}}^2=J_{\text{K}}^a$. 
In \kbets, 
the canted spin structure was observed 
when $\mathbf{H} \parallel c$, 
and we can apply the extended Jaccarino-Peter theory.
However, 
for the magnetic field parallel to the magnetic easy axis, 
i.e., $\mathbf{H} \parallel a$, 
the canted spin structure has not been observed 
and the details of the metamagnetic transition 
have not been revealed.
Therefore, 
we do not apply the present theory to the low-field region 
for $\mathbf{H} \parallel a$.

It is convenient to introduce a constant 
\begin{equation}
\label{eq:pam}
a_m \equiv 
 \left( \frac{\tilde{m}}{m_b} \right)^{1/4}
 \left( \frac{\tvf}{2\pi T_{\text{c}}^{(0)}} \right)^2 
 \frac{2|e|}{c} \;,
\end{equation}
with which $\kappa_x$ and $\kappa_y$ are expressed as
\begin{equation}
\begin{split}
\kappa_x &= a_m H 
 \left( \frac{m_c}{m_a} \right)^{1/4} 
 \left[ \frac{2\pi T_{\text{c}}^{(0)}}{\tvf} \right]^2 
\;, \\ 
 \kappa_y &= a_m H 
 \left( \frac{m_a}{m_c} \right)^{1/4} 
 \left[ \frac{2\pi T_{\text{c}}^{(0)}}{\tvf} \right]^2 \;.
\end{split}
\end{equation}
We also define the strength ratio 
of the paramagnetic and orbital pair-breaking effects 
\begin{equation}
r_m \equiv \frac{h/2\pi T_{\text{c}}^{(0)}}{a_m H} \;.
\end{equation}

We explain the procedure to analyze the experimental phase diagram 
with the present theory.
Our critical field equations contain five microscopic parameters: 
$a_m$, $ m_c / m_a $, $r_m$, $J_{\text{K}}^c$, 
and $J_{\text{K}}^a$. 
First, we fit the experimental data 
of the critical field of the FISC for $\mathbf{H} \parallel c$ 
with the parameters $a_m$, $ m_c / m_a $, $r_m$, 
and $J_{\text{K}}^c$.
Secondly, using the parameter values 
of $a_m$, $ m_c / m_a $ and $r_m$ 
determined in the first step, 
we fit the data of the FISC for $\mathbf{H} \parallel a$ 
with a single parameter $J_{\text{K}}^a$.
It is noted that all five parameters are determined 
only from the curves of the FISC
for $\mathbf{H} \parallel c$ and $\mathbf{H} \parallel a$.
Crudely speaking, the curve of the FISC can be characterized 
by three real numbers, that is, 
the magnetic field where the transition temperature is maximum, 
the maximum transition temperature, 
and the width of the magnetic fields at $T=0$.
Therefore, we obtain six real numbers 
from the two experimental data of the FISC, 
which sufficiently determine 
all five microscopic parameters.
Lastly, using those five parameters determined 
in the first and second steps, 
we calculate the upper critical field 
of low-field superconductivity for $\mathbf{H} \parallel c$, 
where the canted spin structure coexists.
In this last step, 
we do not use any additional fitting parameter.

\begin{figure}[t]
\begin{center}
\includegraphics[width=65mm,keepaspectratio]{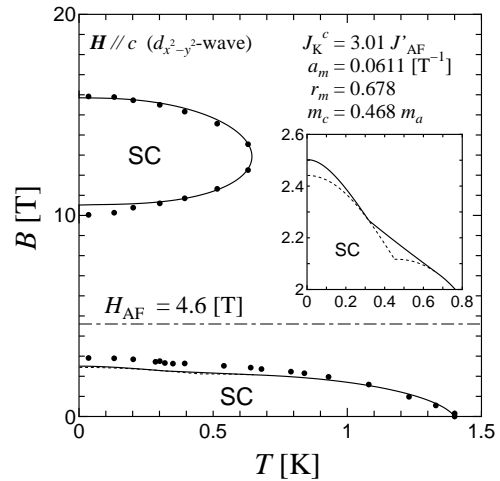}
\caption{\label{fig:d1c}
Theoretical phase diagram for \da-wave 
in the magnetic fields parallel to the \textit{c} axis.
The notation is the same as those in Fig.~\ref{fig:sc}.
The inset shows the theoretical phase diagram below 0.8~K 
and between 2 and 2.6~T.}
\end{center}
\end{figure}

\begin{figure}[htbp]
\begin{center}
\includegraphics[width=65mm,keepaspectratio]{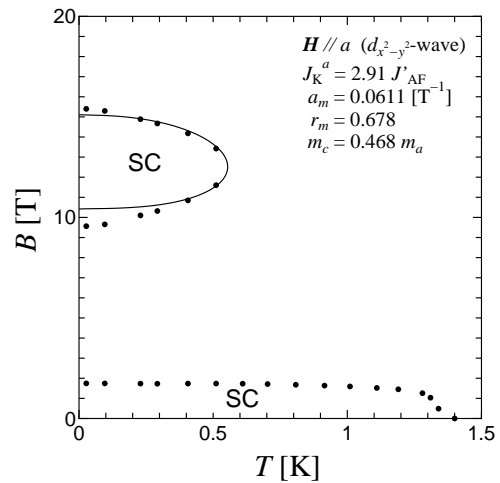}
\caption{\label{fig:d1a} 
Theoretical phase diagram for \da-wave 
in the magnetic fields parallel to the \textit{a} axis.
The notation is the same as those in Fig.~\ref{fig:sc}.}
\end{center}
\end{figure}

\begin{figure}[t]
\begin{center}
\includegraphics[width=65mm,keepaspectratio]{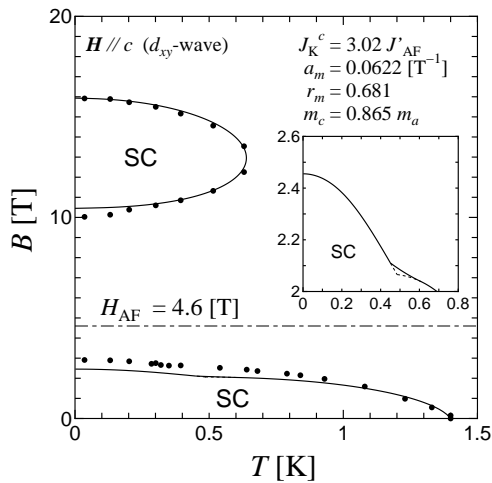}
\caption{\label{fig:d2c}
Theoretical phase diagram for \db-wave 
in the magnetic fields parallel to the \textit{c} axis.
The notation is the same as those in Fig.~\ref{fig:sc}.
The inset shows the theoretical phase diagram below 0.8~K 
and between 2 and 2.6~T.}
\end{center}
\end{figure}

\begin{figure}[t]
\begin{center}
\includegraphics[width=65mm,keepaspectratio]{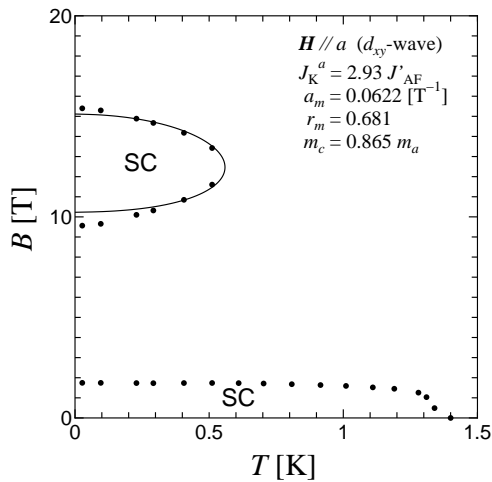}
\caption{\label{fig:d2a} 
Theoretical phase diagram for \db-wave 
in the magnetic fields parallel to the \textit{a} axis.
The notation is the same as those in Fig.~\ref{fig:sc}.}
\end{center}
\end{figure}

\begin{figure}[t]
\begin{center}
\includegraphics[width=65mm,keepaspectratio]{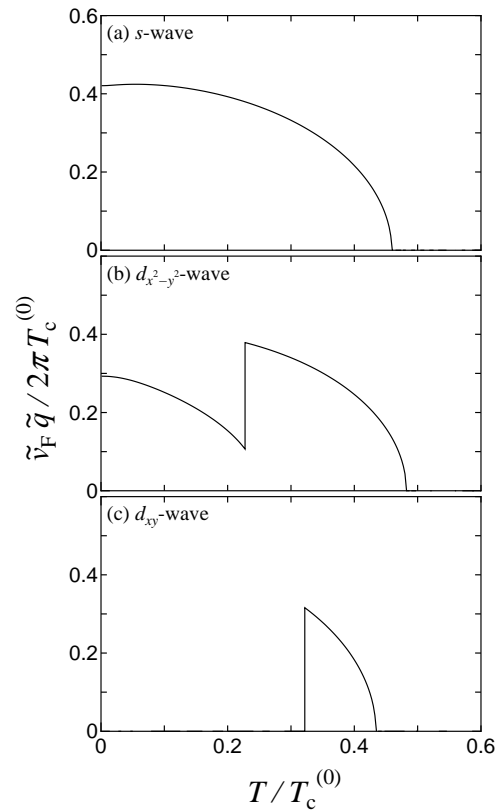}
\caption{\label{fig:ffq} 
Temperature dependence of the magnitudes 
of the total pair momentum 
along the upper critical field curves for 
(a) \textit{s}-wave, (b) \da-wave, and (c) \db-wave pairings 
in the magnetic fields parallel to the \textit{c} axis.
The sets of parameters for (a), (b), and (c) are 
the same as in Figs.~\ref{fig:sc}, \ref{fig:d1c}, and \ref{fig:d2c}, 
respectively.}
\end{center}
\end{figure}

We carry out this procedure for \textit{s}-wave, 
\da-wave, and \db-wave pairings.
In Figs.~\ref{fig:sc}--\ref{fig:ffq}, 
we set $T_{\text{c}}^{(0)}=1.4$~K 
and $H_{\text{AF}}=4.6$~T, 
which have been observed in the compounds \kbets.
For all pairing symmetries examined, 
the theoretical curves agree well with the experimental data, 
as shown in Figs.~\ref{fig:sc}--\ref{fig:d2a}. 
In particular, 
the agreements at low field for $\mathbf{H} \parallel c$ 
are quantitative, 
because all parameters are determined only from the data of the FISC 
at high fields.
In what follows, 
we explain the differences depending on the pairing symmetries.

The results for \textit{s}-wave pairing 
are shown in Figs.~\ref{fig:sc} and \ref{fig:sa}.
We obtain the effective-mass ratio $m_c/m_a=0.785$, 
which is reasonable in comparison with the value estimated 
from the Shubnikov--de Haas~(SdH) oscillations~\cite{konoike05,balicas00}. 
It is confirmed that the FFLO state is not realized in the FISC. 
In contrast, it is shown that low-field superconductivity 
exhibits the FFLO state at low temperatures.

The results for \da-wave pairing 
are shown in Figs.~\ref{fig:d1c} and \ref{fig:d1a}.
We obtain $m_c/m_a=0.468$, 
which is too small in comparison with the value estimated 
from the SdH oscillations~\cite{konoike05,balicas00}.
It is confirmed that the FFLO state is not realized in the FISC. 
There are  bends in the theoretical curves of the low fields, 
where the Landau level indexes $n$'s of the optimum solution alternate.
The solution near $T=0$ is expressed 
by a linear combination of the Abrikosov functions 
with odd $n$'s.
In Fig.~\ref{fig:d1c}, 
it is found that there are two tricritical points 
in the phase diagram for $\mathbf{H} \parallel c$.
One at $T \simeq 0.67$~K is the critical point of the normal state, 
the BCS state with even $n$'s, 
and the FFLO state with even $n$'s, 
and another at $T \simeq 0.32$~K is that of the normal state, 
the FFLO states with even $n$'s and odd $n$'s.

The results for \db-wave pairing 
are shown in Figs.~\ref{fig:d2c} and \ref{fig:d2a}.
We obtain $m_c/m_a=0.865$, 
which is reasonable in comparison with the value estimated 
from the SdH oscillations~\cite{konoike05,balicas00}.
In this case, the area of the FFLO state is very limited.
It is found that the FFLO state is not realized in the FISC 
and low-field superconductivity near $T=0$. 
There is a bend in the theoretical curve, 
where the parity of the Landau-level indexes $n$'s alternates.
The solution of the odd $n$'s gives the highest critical field near $T=0$.

In Fig.~\ref{fig:ffq}, 
we show temperature dependences of 
$\tilde{q} \equiv (\tilde{m}/m_c)^{1/2}|\bq|$ 
along the upper critical field curves 
of low-field superconductivity for 
\textit{s}-wave, \da-wave, and \db-wave pairings 
when $\mathbf{H} \parallel c$.
For \da-wave pairing, 
it is found that the $\tilde{q}$ jumps 
at a temperature where the Landau-level indexes $n$ 
of the optimum solution alternate.
For \db-wave pairing, 
it is found that the $\tilde{q}$  
vanishes for $T < 0.32 T_{\text{c}}^{(0)}$, 
where the FFLO state with odd $n$ is unstable.

In a brief summary of this section, 
experimental results could be reproduced 
for all \textit{s}-wave, \da-wave, and \db-wave parings. 
For $s$-wave pairing, 
the upper critical fields of low-field superconductivity 
are smaller than the experimental data, 
but they are improved by taking into account the FFLO state. 
The in-plane mass anisotropies are estimated as $m_c/ m_a = 0.785$, 
0.468, and 0.865 for \textit{s}-wave, \da-wave, and \db-wave pairings, respectively.
The results for \textit{s}-wave and \db-wave pairings agree 
with the value obtained from the SdH oscillations~\cite{konoike05,balicas00}, 
whereas the result for \da-wave pairing largely deviates from it.
We found that the FFLO state does not occur in the FISC, 
while it occurs in low-field superconductivity.

\section{\label{sec:sd}Summary and discussion}

We have studied the upper critical fields 
of antiferromagnetic \textit{s}-wave, 
\da-wave, and \db-wave superconductors 
with Fermi surface anisotropy.
We have derived the linearized gap equations 
when the magnetic field is applied parallel to layers, 
taking into account the effective-mass anisotropy, the FFLO state, 
and the extended Jaccarino-Peter mechanism. 
We have applied the theory 
to the organic superconductor \kbets, 
and obtained good agreement between the theoretical 
and experimental results.

\begin{figure}[t]
\begin{center}
\includegraphics[width=65mm,keepaspectratio]{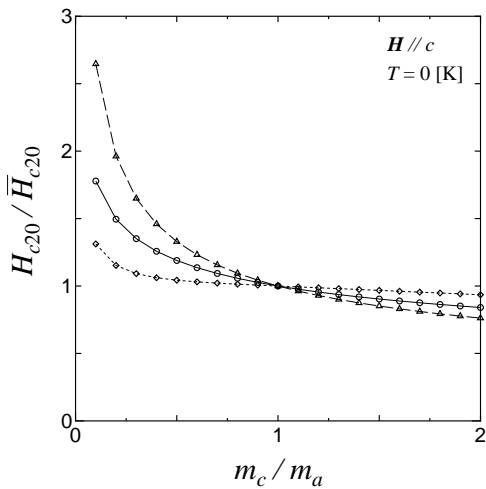}
\caption{\label{fig:mae} The mass anisotropy dependence 
of the pure orbital limits for $\mathbf{H} \parallel c$.
The open circles, open diamonds, and open triangles show 
the results for \textit{s}-wave, \da-wave, and \db-wave pairings, 
respectively.
The solid, dotted, and dashed curves are guides to the eye. 
Here, $\bar{H}_{c20}$ denotes the orbital limit $H_{c20}$ 
of the isotropic case ($m_c=m_a$).}
\end{center}
\end{figure}

While the agreements have been obtained 
for all cases of \textit{s}-wave, \da-wave, and \db-wave pairings, 
the resultant in-plane mass ratios $m_c / m_a$ are different from one another.
The mass ratio for \da-wave pairing seems too small, 
although those for \textit{s}-wave and \db-wave pairings are 
close to the value estimated 
from the SdH oscillations~\cite{konoike05,balicas00}.
The mass anisotropy dependence of the pure orbital limits 
of the upper critical fields is shown in Fig.~\ref{fig:mae}.
The upper critical fields for 
\textit{s}-wave, \da-wave, and \db-wave pairings 
become larger when the ratio $m_c/m_a$ decreases.
It is found that the upper critical field 
is the most influenced for \db-wave pairing 
by a change in the ratio $m_c/m_a$, 
while it is less influenced for \da-wave pairing. 
This is why the resultant ratio $m_c/m_a$ 
to fit the experimental data becomes much smaller 
for \da-wave pairing than for \db-wave pairing.

This result is physically interpreted as follows.
For example, we consider the situation that $m_c/m_a \le 1$.
The origin of the orbital pair-breaking effect 
is the Lorentz force that acts on electrons.
When the magnetic fields are applied along the \textit{c} axis, 
the Lorentz force is the strongest for electrons 
near the $p_a$ axis, where the Fermi velocity is the largest, 
in the two-dimensional momentum space.
When the mass ratio $m_c/m_a$ decreases with the fixed electron number, 
i.e., when the effective mass $m_a$ increases, 
the Fermi velocity in the \textit{a} direction decreases. 
Therefore, the Lorentz force acting on 
the electrons on the Fermi surface near the $p_a$ axis 
becomes stronger and superconductivity is more suppressed 
when $m_c/m_a$ decreases.
The order parameter of \db-wave pairing 
has nodes on the $p_a$ axis, 
while that of \da-wave pairing has the largest amplitude there.
Therefore, \da-wave pairing is less affected 
than \db-wave pairing by a change in the ratio $m_c/m_a$.

It has been found from Fig.~\ref{fig:d2c} 
that the FFLO state does not occur for \db-wave pairing 
in low-field superconductivity near $T=0$.
This result is physically explained as follows.
Since the effective density of states for $\alpha$-wave pairing 
is proportional to $[\gamma_{\alpha}(\hat{\bp})]^2$ in the gap equation, 
areas of the Fermi surface near the peak of $[\gamma_{\alpha}(\hat{\bp})]^2$ 
have a greater effect on the nesting condition for the FFLO state, 
but those near the nodes of $[\gamma_{\alpha}(\hat{\bp})]^2$ 
have less effect~\cite{shimahara97b}. 
Hence, 
in the absence of the orbital pair-breaking effect, 
the vector $\bq$ tends to points in the areas of the Fermi surfaces 
where $[\gamma_{\alpha}(\hat{\bp})]^2$ exhibits a peak. 
In this study, however, we adopt the model 
in which $\bq \parallel \mathbf{H}$ is assumed 
due to the orbital pair-breaking effect. 
Therefore, for \db-wave pairing, 
the Fermi surface nesting is less effective 
when $\mathbf{H} \parallel {\hat x}$ or ${\hat y}$, 
because the order parameter has nodes on the $p_x$ and $p_y$ axes. 
In contrast, 
for \da-wave pairing, 
because the order parameter has peaks 
on the $p_x$ and $p_y$ axes, 
the FFLO state occurs in low-field superconductivity, 
when $\mathbf{H} \parallel {\hat x}$ or ${\hat y}$.

\begin{figure}[t]
\begin{center}
\includegraphics[width=65mm,keepaspectratio]{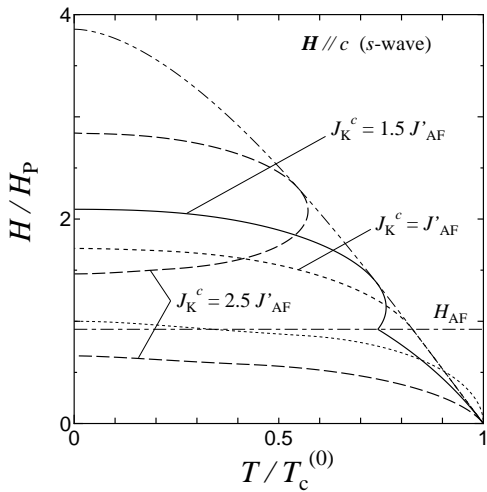}
\caption{\label{fig:sj} 
Theoretical predictions of the phase boundary of superconductivity 
for various values of $J_{\text{K}}^c $. 
The other parameters are common to Fig.~\ref{fig:sc}. 
The thin dotted and thin two-dot-dashed curves show the Pauli paramagnetic 
and orbital limits, respectively.
In this figure, 
the FFLO state is taken into account, 
but modifies the curves slightly at low temperatures. 
At the upper critical fields of the FISC 
for $J_{\text{K}}^c = 2.5 J_{\text{AF}}' $, 
the FFLO state does not occur.}
\end{center}
\end{figure}

In the application to \kbets, it is found that 
the FFLO state is not present in the FISC 
in contrast to \lbets, 
while it is present in low-field superconductivity. 
In \lbets, the phase boundary of the FISC 
can be well reproduced if the possibility of the FFLO state 
is taken into account~\cite{shimahara02b}. 
For the FFLO state to occur, 
the orbital pair-breaking effect needs to be sufficiently suppressed.
In \lbets, this condition seems to be satisfied 
even at high field due to low dimensionality.
In contrast, 
it was suggested that \kbets\ 
has a larger interlayer electron hopping energy than \lbets~\cite{konoike04}.
Therefore, in \kbets, 
the orbital effect is substantial at high fields, 
as verified by comparison of the zero-field transition temperature 
and the maximum transition temperature of the FISC.

The upper critical fields of low-field superconductivity 
agree with the experimental data, 
as shown in Figs.~\ref{fig:sc}, \ref{fig:d1c}, and \ref{fig:d2c}. 
Further, for $s$-wave pairing, 
the agreement is improved if the FFLO state is taken into account. 
However, 
they are slightly smaller than the experimental data. 
This discrepancy may be removed 
by taking into account a mixing effect 
in the presence of the weak triplet pairing interaction, 
as in the compounds \lbets~\cite{shimahara02b}. 
A slight discrepancy also occurs in the lower critical fields of the FISC, 
where the FFLO state does not occur as explained above. 
If the mixing effect occurs, this discrepancy may also be removed.

For \da-wave and \db-wave pairings in low-field superconductivity, 
we found an internal transition 
between the vortex states with different Landau level indexes $n$, 
which is analogous to internal transitions in a two-dimensional system 
in tilted magnetic fields~\cite{shimahara97a}. 
Both the FFLO state 
and the vortex state with higher $n$'s 
originate from the Pauli paramagnetic effect 
as discussed in Ref.~\cite{shimahara97a}, 
because in the absence of Zeeman energy 
only the vortex state with lower $n$'s occurs. 
Due to the transitions to 
the FFLO state and 
the vortex states with higher $n$'s, 
the upper critical fields of low field superconductivity 
exhibit downward convex curves for all pairing symmetries examined, 
which agrees with the experimental data.

We have determined the microscopic parameters only 
from the curves of the FISC, 
and calculated the upper critical field 
of low-field superconductivity with those parameter values 
without any additional fitting parameters.
From the accordance between the theoretical and experimental results 
on low-field superconductivity, 
the extended Jaccarino-Peter mechanism~\cite{shimahara02a,shimahara04} seems 
to be realized in \kbets.
Applying the extended Jaccarino-Peter mechanism 
to the same models with different parameter values, 
we obtain some unusual phase diagrams as shown in Fig.~\ref{fig:sj}. 
When $J_{\text{K}}^c = 1.5 J_{\text{AF}}'$, 
the transition temperature curve has double peaks. 
When $J_{\text{K}}^c = J_{\text{AF}}'$,
superconductivity occupies large area in the phase diagram. 
Similar phase diagrams have been obtained 
in two- and three-dimensional systems~\cite{shimahara02b,shimahara04}. 
In this study, 
we confirmed that the phase diagrams as obtained above 
can be realized in a more realistic model of organic superconductors. 
If we control the value of $J_{\text{K}}$ by material design, 
such critical field curves would be observed.

In the present theory, 
the effective mass $m_b$ is included in the parameter $a_m$. 
Since we have obtained the value of $a_m$, 
we can estimate a value of $m_b$, 
if we have the value of $\tp_{\text{F}}$ appropriate for \kbets.

The upper critical field in low fields for $\mathbf{H} \parallel a$ 
could not be reproduced by the present theory.
It has been found that 
the magnetic easy axis of antiferromagnetic long-range order in low fields 
is along the \textit{a} axis~\cite{fujiwara01}, 
and that the metamagnetic transition occurs 
at the antiferromagnetic phase boundary close to 
the upper critical field~\cite{konoike04}.
In order to reproduce the upper critical field 
in low fields for $\mathbf{H} \parallel a$, 
we need to reproduce the localized spin structure.
This remains for future study.

In conclusion, the experimental phase diagrams of the antiferromagnetic 
superconductor \kbets\ are theoretically reproduced 
by the present theory.
In particular, 
the low-field phase for $\mathbf{H} \parallel c$ 
is well reproduced by the model parameters that 
are determined from the critical fields of the FISC.
Therefore, the extended Jaccarino-Peter mechanism 
seems to be realized in the present compound at low fields.
Due to this mechanism, 
some unusual phase diagrams 
may occur in compounds with appropriate energy parameters, 
in the future.

\begin{acknowledgments}

We would like to thank 
S.~Uji and T.~Konoike for discussions 
and for providing us with the experimental data.
We wish to acknowledge useful discussions with K.~Doi.
This work was partially supported by the Ministry of Education, Science,
Sports and Culture of Japan, Grant-in-Aid for Scientific
Research~(C), No.~16540320, 2005.

\end{acknowledgments}

\appendix



\end{document}